\documentclass[pra,twocolumn,showpacs,amsmath,floatfix]{revtex4}
\usepackage{graphicx}
\usepackage{amssymb}

\begin{document}

\def\ket#1{|#1\rangle}
\def\bra#1{\langle#1|}
\def\av#1{\langle#1\rangle}
\def\myarrow{\mathop{\longrightarrow}}

\title{Two-level atom excitation probability for single- and $N$-photon wavepackets}

\author{Hemlin Swaran Rag and Julio Gea-Banacloche}
\affiliation{Department of Physics, University of Arkansas, Fayetteville, AR 72701}
\email[]{jgeabana@uark.edu}

\date{\today}

\begin{abstract}
We study how the transient excitation probability of a two-level atom by a quantized field depends on the temporal profile of the incident pulse, in the presence of external losses, for both coherent and Fock states, and in two complementary limits: when the pulse contains only one photon (on average), and when the number of photons $N$ is large.  For the latter case we derive analytical expressions for the scaling of the excitation probability with $N$ that can be easily evaluated for any pulse shape.  
\end{abstract}


\maketitle

\section{Introduction and motivation}
Unlike its steady-state counterpart, the transient excitation probability of a single two-level atom interacting with a quantized field can, in principle, approach unity, for times smaller than the excited-state lifetime.  Such perfect or near-perfect excitation could be useful in, for instance, quantum information processing, as a way to implement a single-atom switch, or a logical gate.  The case in which the field consists of a single photon, in particular, has generated a fair amount of interest over the years.  Schemes involving one-dimensional waveguides \cite{shenfan1,scarani1,chen} as well as free-space interaction \cite{leuchs1,leuchs2,leuchs3} have been considered recently.

An important result from these studies is the realization that the excitation probability depends critically, not just on the photon pulse's transverse-mode profile (in the case of free-space excitation), but also on its temporal profile.  In particular, it has been known for a long time \cite{leuchs2} that the only way to achieve unit excitation probability is to use a wavepacket that is the time-reversed version of the one emitted by the atom when it decays spontaneously, which is to say, in free space, a ``rising exponential'' pulse (the notion of using a time-reversed pulse was first introduced, to our knowledge, in the context of cavity QED \cite{cirac1}).  A number of schemes to generate such pulses have been proposed and partly demonstrated in recent years \cite{du1,leuchs4,leuchs5,leuchs6,martinis,du2,kurtsiefer1,kurtsiefer2}.

Our goal in this paper is twofold.  In the first part (Section II), we use the model developed in \cite{scarani1,scarani2} to study theoretically the excitation probability for a two-level atom by a single-photon pulse, as a function of the temporal profile of the pulse, in the presence of external losses.  These ``losses'' can be used to model what happens when the coupling to the spatial profile of the incident pulse is not perfect, that is to say, the atom is coupled to, and can decay into, other spatial field modes.  Our results are therefore quite general and can apply both to the waveguide and free-space configurations.  Besides the inclusion of losses, we also consider some novel temporal profiles, and generally carry our analytical calculations a bit farther than most previous studies, although in the end the final optimization of the pulse bandwidth typically needs to be done numerically.  

In the second part (Section III), we turn our attention to the problem of excitation by multiphoton pulses, again in the presence of external losses and for various temporal profiles, and explore how the maximum excitation probability scales asymptotically with the number of photons in the pulse.  The motivation for this comes initially from a consideration of the minimum energy requirements for quantum logic \cite{geabanaPRL}.  It also complements the research reported in the first part, inasmuch as some schemes to generate single-photon rising exponential pulses may involve filtering, or otherwise throwing away a potentially large number of photons, and may only approximately succeed at generating the required shape.  At that point, it makes sense to explore the asymptotic behavior of the excitation probability to ascertain whether one might not more efficiently resort to direct excitation of the atom by a more conventional, multi-photon pulse.  (Of course, one does not have that luxury when the single photon is itself a qubit, or carrier of quantum information, but this does not always need to be the case.)

Throughout the paper, we consider both multimode Fock states and coherent states, for completeness, although in practice Fock states make more sense in the context of single-photon pulses, and coherent states in the context of multiphoton pulses. For the latter, we will present an analytical treatment that makes it straightforward to calculate the asymptotic (large $\bar n$), optimized excitation probability, for an arbitrary pulse shape.   

\section{Single-photon results}

\subsection{Fock states} 

\subsubsection{General equations, and rising exponential pulse}

For a single-photon pulse in a state of arbitrary temporal profile $f(t)$ (assuming $\int_{-\infty}^\infty |f(t)|^2\, dt = 1$), the on-resonance equations for the excitation of a two-level atom are:
\begin{align}
\dot P_e &= -(\Gamma_P+ \Gamma_B) P_e - \sqrt{\Gamma_P} f(t) \Sigma \cr
\dot \Sigma &= -\frac 1 2(\Gamma_P+ \Gamma_B) \Sigma - 2\sqrt{\Gamma_P} f(t) 
\label{e1}
\end{align}
Here we denote by $\Gamma_P$ the coupling to the spatial modes that make up the incoming pulse, and by $\Gamma_B$ the coupling to other, ``bath'' modes (which results in loss of a photon from the system). $P_e$ is the excitation probability, and $\Sigma$ is the matrix element of the atomic dipole moment in between the states with 1 and 0 photons.  For a derivation of these equations, see \cite{scarani1,domokos} (note that our $\Sigma$ is the sum of the $\sigma_+$ and $\sigma_-$ variables in \cite{scarani1}; also, we are assuming $f(t)$ is real, which is a natural assumption on resonance).

The system (\ref{e1}) is simple enough to allow for a general, formal solution for arbitrary $f(t)$: the equation for $\Sigma$ can be immediately integrated, and then we have for $P_e$ (assuming the atom starts in the ground state)
\begin{align}
P_e(t) = 2\Gamma_P &\int_{-\infty}^t e^{-(\Gamma_P+ \Gamma_B)(t-t^\prime)}f(t^\prime)\, dt^\prime \cr
&\times\int_{-\infty}^{t^\prime}  e^{-(\Gamma_P+ \Gamma_B)(t^\prime-t^{\prime\prime})/2}f(t^{\prime\prime})\, dt^{\prime\prime} 
\label{e2a}
\end{align}
Inspection (or an integration by parts) shows that this can be rewritten in the alternative form
\begin{equation}
P_e(t) = \Gamma_P e^{-(\Gamma_P+\Gamma_B)t} \left(\int_{-\infty}^t e^{(\Gamma_P+\Gamma_B)t^\prime/2} f(t^\prime)\, dt^\prime\right)^2
\label{e2}
\end{equation}
which makes the calculation much simpler, for arbitrary-shaped wavepackets.  Equation (\ref{e2}) also allows for a very simple proof that the only wavepacket that can achieve full excitation at any time is a rising exponential (see \cite{leuchs2,shenfan1} for alternative approaches), in the absence of external losses.  Consider $P_e$ at an arbitrary time $t=t_0$.  We can rewrite Eq.~(\ref{e2}) as
\begin{equation}
P_e(t_0) = \frac{\Gamma_P}{\Gamma_P+\Gamma_B} \left(\int_{-\infty}^{t_0} u(t) f(t) \, dt \right)^2
\label{e4a}
\end{equation}
where the function $u(t)$, defined as
\begin{equation}
u(t) = {\sqrt{\Gamma_P+\Gamma_B}}\, e^{(\Gamma_P+\Gamma_B)(t-t_0)/2}
\end{equation}
is normalized to unity in the interval $(-\infty,t_0]$.  From the Cauchy-Schwartz inequality, it then follows immediately that
\begin{equation}
P_e(t_0) \le  \frac{\Gamma_P}{\Gamma_P+\Gamma_B} \int_{-\infty}^{t_0}  f(t)^2 \, dt 
\label{e6a}
\end{equation}
However, since $f(t)$ is normalized to unity in $(-\infty,\infty)$, it follows that the right-hand side of (\ref{e6a}) can never be equal to 1 unless, first, $\Gamma_B=0$ (no external losses) and, secondly, all of the norm of $f$ is contained in $(-\infty,t_0]$ (otherwise put, the pulse must be over by the time $t_0$).  But then, the integral in (\ref{e4a}) is just the inner product of two functions normalized to unity over the interval $(-\infty,t_0]$, and so it can only be equal to 1 (its maximum possible value) if the functions are identical except for an overall sign.  

We conclude, then, that $P_e(t_0)$ can only reach a maximum value of 
\begin{equation}
P_{e,max} = \frac{\Gamma_P}{\Gamma_P + \Gamma_B}  \qquad \text{(rising exponential)}
\label{opt}
\end{equation}
at some time $t_0$, if the excitation pulse has the form
\begin{equation}
f(t) = {\sqrt{\Gamma_P+\Gamma_B}}\, e^{(\Gamma_P+\Gamma_B)(t-t_0)/2}, \qquad \text{$t\le t_0$, 0 if $t>t_0$}
\label{n8}
\end{equation}
When $\Gamma_B =0$, we have $P_{e,max}=1.$  

Note that, in general, in the presence of external losses, the optimal duration (bandwidth) of the pulse needs to be adjusted to include the $\Gamma_B$ term (as in Eq.~(\ref{n8})). When this is done, it is evident from Eq.~(\ref{e4a}) that the maximum excitation probability will be found to be equal to the lossless result times the factor $\Gamma_P/(\Gamma_P+\Gamma_B)$.  

It is a relatively straightforward matter to use Eq.~(\ref{e2}), or equivalently (\ref{e4a}), to derive the excitation probability for other pulse shapes, to see how close they may get to the optimal result (\ref{opt}).  We present several of these results explicitly below.  (Some of these, in the lossless case, were previously presented in \cite{scarani1}, where they appear to have been obtained by numerical integration of the equations (\ref{e1}).  This had, in particular, the curious consequence that the maximum excitation probability reported for the optimal rising exponential pulse was $0.995$ instead of 1.) 

\subsubsection{Square pulse}
For a square pulse of duration $T$: $f(t) = 1/\sqrt T$, $0<t<T$, the first Eq.~(\ref{e1}) shows that $P_e$ starts to decay as soon as the pulse is over, so to find the maximum we may confine ourselves to the region $0<t<T$, in which case we get from Eq.~(\ref{e2}),
\begin{equation}
P_e(t) = \frac{4 \Gamma_P}{(\Gamma_B+\Gamma_P)^2 T} \left(1-e^{-(\Gamma_B+\Gamma_P)t/2}\right)^2
\end{equation}
This is maximum for $t=T$, and then we can optimize for $T$.  Numerically we find $T_\text{opt} = 2.513/(\Gamma_B+\Gamma_P)$, and so
\begin{equation}
P_{e,max} = 0.815\frac{\Gamma_P}{\Gamma_P + \Gamma_B} \qquad \text{(square pulse)}
\end{equation}

\subsubsection{Gaussian pulse}
If we consider a Gaussian pulse instead, of the form 
$f(t) = e^{-t^2/T^2}/\sqrt{T\sqrt{\pi/2}}$, substitution in (\ref{e2}) yields the exact expression
\begin{align}
P_e(t) = &\frac{\sqrt{2\pi} \Gamma_P T}{4} e^{-(\Gamma_P+\Gamma_B)t + (\Gamma_P+\Gamma_B)^2 T^2/8}\cr
&\times\left[1+\text{erf}\left(\frac t T - \frac{(\Gamma_P+\Gamma_B) T}{4}\right)\right]^2
\end{align}
Numerical maximization of this expression with respect to $t$ and $T$ yields $t_\text{opt}  \simeq 0.731 T$, $T_\text{opt}  = 1.368/(\Gamma_P+\Gamma_B)$, and
\begin{equation}
P_{e,max} = 0.801 \frac{\Gamma_P}{\Gamma_P + \Gamma_B} \qquad \text{(Gaussian)}
\end{equation}
It is interesting that the performance of the Gaussian pulse is extremely close to that of the square pulse.  In the next section we will see that this is the case in the multiphoton, asymptotic limit as well.

\subsubsection{Pulses obtained by atomic decay}
We next look at a couple of pulses that might be easier to produce experimentally than the ones considered above.  One of these is a simple exponentially-decaying pulse,  $f(t) = e^{-t/T}\sqrt{2/T}$ for $t\ge 0$ (and zero for $t<0$).  Equation (\ref{e2}) now yields
\begin{equation}
P_e(t) = \frac{8\Gamma_P T}{(\Gamma_P T + \Gamma_B T-2)^2}\left(e^{-t/T}-e^{-(\Gamma_B+\Gamma_P)t/2}\right)^2
\label{e7}
\end{equation}
As a function of $t$, this expression peaks at 
\begin{equation}
t_{max} = \frac{2T}{\Gamma_P T + \Gamma_B T-2}\,\ln\left[(\Gamma_P + \Gamma_B )T/2\right]
\label{e8}
\end{equation}
Substitution of this back into Eq.~(\ref{e7}) leads to a complicated expression which, however, can be shown to have a maximum, as a function of $T$, when $T = 2/(\Gamma_P + \Gamma_B)$ (in which limit the expression (\ref{e8}) becomes $t_{max} = T$).  This maximum value equals
\begin{align}
P_{e,max} &= \frac{4}{e^2}\, \frac{\Gamma_P}{\Gamma_P + \Gamma_B} \quad \; \text{(decaying exponential)}\cr
&\simeq 0.541 \frac{\Gamma_P}{\Gamma_P + \Gamma_B} 
\end{align}
A somewhat more complex, but still, experimentally, relatively straightforward, kind of pulse would be the one produced by an atom decaying inside a single-sided cavity.  If the atom is assumed to be fully excited at the time $t=0$, the pulse for $t\ge 0$ is given by 
$f(t)=-\frac{g\sqrt{2\kappa}}{\sqrt{\kappa^2-4g^2}}\left(e^{-\left(\kappa+\sqrt{\kappa^2-4g^2}\right)t/2}-e^{-\left(\kappa-\sqrt{\kappa^2-4g^2}\right)t/2}\right)$ 
(see \cite{gea1}, Eq.(54)), where $g$ is the coupling rate of the atom to the cavity, and $\kappa$ the cavity decay rate. The excitation probability with such a pulse is
\begin{align}
P_e(t) = &\dfrac{8g^2\kappa\Gamma_{p}\,e^{-(\Gamma_{p}+\Gamma_{B})t}}{\kappa^2-4g^2}\Biggl(\dfrac{e^{\left(\Gamma_{p}+\Gamma_{B}-\kappa+\sqrt{\kappa^2-4g^2}\right)t/2}-1}{\Gamma_{p}+\Gamma_{B}-\kappa+\sqrt{\kappa^2-4g^2}}\cr
&-\dfrac{e^{\left(\Gamma_{p}+\Gamma_{B}-\kappa+\sqrt{\kappa^2-4g^2}\right)t/2}-1}{\Gamma_{p}+\Gamma_{B}-\kappa-\sqrt{\kappa^2-4g^2}}\Biggr)^2
\label{ne16}
\end{align}
We have not been able to find the maximum of this expression (with respect to all three parameters, $\kappa$, $g$ and $t$) analytically.  Numerically, however, we have found that the optimal pulse happens in the good cavity limit, that is $\kappa < 2g$, so the square roots in (\ref{ne16}) are purely imaginary, and the time dependance includes Rabi oscillations as well as exponential decay.  We have also found numerically that, in this region, the optimal value of $\kappa$ is given by $\kappa=\Gamma_{p}+\Gamma_B$, in a similar way as for the simple decaying exponential.  This observation allows us to solve for the optimal time, with the result 
\begin{equation}
t_{max}=
\frac{4}{\sqrt{4g^2-\kappa^2}}\,\tan^{-1}\frac{\sqrt{4g^2-\kappa^2}}{\kappa}
\end{equation}
The final maximization with respect to $g$ has to be done again numerically, with the result $g_{opt} = 0.9076 \kappa = 0.9076(\Gamma_P+\Gamma_B)$ (which means $t_{max} = 2.607/(\Gamma_P+\Gamma_B))$, and
\begin{equation}
P_{e,max} = 0.716 \frac{\Gamma_P}{\Gamma_P + \Gamma_B} \qquad \text{(Atom-cavity decay pulse)}
\end{equation}
Thus, in spite of all the extra parameters, this family of pulses still cannot do better than the Gaussian or the square, although it is certainly better than the plain decaying exponential.  

The atom-cavity system, however, could in principle be used to generate a much greater variety of pulses, depending on how it is driven, itself. Thus, for instance, one could think of sending a single-photon pulse (with a simple shape, such as a Gaussian or a decaying exponential) into the cavity, through the coupling mirror, and then using the output pulse to excite the target atom in the waveguide.  The output pulse profile is, for any input pulse, easily derived from the results in \cite{gea1}.  Our calculations show that the efficiency of an initial Gaussian pulse can be boosted in this way to $0.85$, for example.  (See the second line from the top in Fig. 1, which summarizes all the above results graphically.)

\begin{figure}
\begin{center}
\includegraphics[width=3.5in]{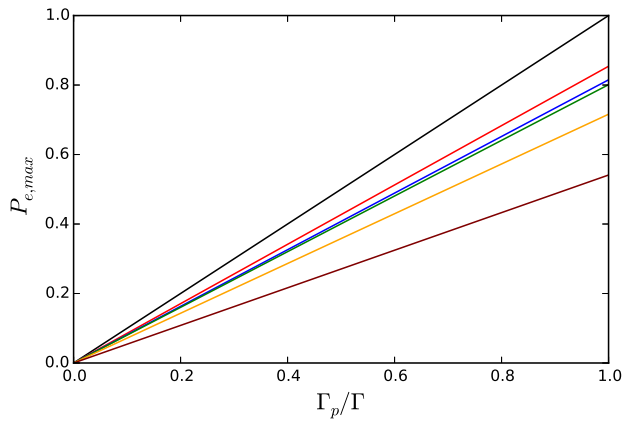}
\end{center}
\caption[example]
   { \label{fig:fig1}
Optimized excitation probability for various single-photon pulse shapes, as a function of the ratio of coupling $\Gamma_P$ to coupling plus losses, $\Gamma = \Gamma_P+\Gamma_B$.  From top to bottom, the lines are for a rising exponential pulse, a Gaussian pulse modified by interacting with an atom in a single-sided cavity, a ``square'' pulse, an ordinary Gaussian pulse, a pulse emitted by an (initially excited) atom in a cavity, and a decaying exponential pulse.  }
\end{figure}

Another possibility would be to drive the atom in the cavity directly (through the sides of the cavity, say), and near-deterministically, with a sufficiently strong external field.  By controlling the time dependence of the atomic excitation in this way, one could in principle control the shape of the outgoing pulse (which would still be a single photon pulse, as long as care is taken not to cycle the atomic excitation up and down more than once) \cite{zoller}.  Note, however, that at this point we are talking about using many photons (in the control field) just to generate a single-photon wavepacket with the ideal profile to perfectly excite a single atom.  In some contexts, as when one means to use the single photon as a qubit, this may make sense; but if all we want is to excite a single atom with a minimal expenditure of energy, it seems reasonable to try a different tack and ask instead just how many photons, impinging directly on the target atom, it would take to bring its excitation probability arbitrarily close to one, assuming either that one starts with a pulse with the ``wrong'' shape (i.e., not a rising exponential), or that the coupling losses to the outside world represented by $\Gamma_B$ are not negligible.  

This is the question that we will address in the second part of this paper, after we briefly consider the atomic excitation by ``single-photon'' coherent state wavepackets of various shapes in the next subsection.

\subsection{Coherent states}

As shown in \cite{scarani1,scarani2}, if the incident pulse is in a coherent state instead of a number state, the quantized-field treatment yields a result formally identical to the semiclassical ``optical Bloch equations.'' If the field is on resonance, the atom initially in the ground state, and the average number of incident photons is $\bar n$, then the atomic dipole moment and excitation probability are given by 
\begin{align}
\begin{split}
& \dot{\Sigma} =-\dfrac{\Gamma_{B}+\Gamma_{p}}{2}\Sigma +4\sqrt{\bar{n}\Gamma_{p} }f(t)P_{e}-2\sqrt{\bar{n}\Gamma_{p} }f(t)
\cr
& \dot{P_{e}} =-(\Gamma_{B}+\Gamma_{p}) P_{e} -\Sigma f(t)\sqrt{\bar{n}\Gamma_{p} }
\end{split}
\label{ne19}
\end{align}
where, as before, $f(t)$ is the pulse profile.  In the absence of damping ($\Gamma_P+\Gamma_B=0$), these equations are easy to solve, and lead to the familiar result that full inversion is achieved by using a $\pi$ pulse, that is to say, one for which $f$ satisfies
\begin{equation}
2 \sqrt{\bar n \Gamma_P} \int f(t) \, dt = \pi
\label{ne20}
\end{equation}
Of course, because of the presence of $\Gamma_P$ in (\ref{ne20}), this condition is, strictly speaking, incompatible with the setting of $\Gamma_B+\Gamma_P =0$, but this might still be approximately valid for a sufficiently intense ($\bar n \gg 1$) and short ($(\Gamma_B+\Gamma_P)T \ll 1$) pulse.  This will be discussed further in Section III.

For this section, we only want to consider the case of a ``single-photon'' coherent-state pulse.  By this we mean a pulse with $\bar n =1$.  When expressed in terms of Fock states, such a pulse has a probability $p(n) = e^{-1}/n!$ to contain $n$ photons, that is to say, a probability $1/e = 0.368\ldots$ of having 0 photons (in which case no excitation will happen), an identical probability of having 1 photon, and a probability $1-2/e = 0.264\ldots$ of having more than 1 photon.  Therefore, in terms of the single-photon Fock state excitation probability discussed in the previous subsection, which we will call $P_{e,N=1}$ below, we can bound the coherent-state excitation probability $P_{e,\bar n =1}$ by
\begin{equation}
0.368 P_{e,N=1} < P_{e,\bar n =1} < 0.632
\label{ne21}
\end{equation}
for any pulse shape.  (The upper limit is just $1-p(0)$.)  

Equation (\ref{ne21}) is enough to see that ``single-photon'' coherent state pulses can never achieve very large excitation probabilities, regardless of their shape.  Numerical results for these pulses have been presented in \cite{scarani1}.  Here we will only consider the one analytically solvable case, the square pulse, because we can do it for arbitrary $\bar n$, and the large $\bar n$ limit will be useful in the next section.  Letting, then, $f(t) = 1/\sqrt T$, $0<t<T$, and defining $\Gamma = \Gamma_P+\Gamma_B$ for simplicity, $\Omega_0=2\sqrt{\bar{n}\Gamma_{p}/T}$, and $\Omega=\sqrt{\Omega^2-\Gamma^2/16}$, we get
\begin{equation}
P_{e}=\dfrac{\Omega_0^2}{\Gamma^2+2\Omega_0^2}\left(1-e^{-3/4~\Gamma t }\left[\cos(\Omega t)+\dfrac{3\;\Gamma}{4 \Omega}\sin(\Omega t)\right]  \right)
\label{ne22}
\end{equation}
Maximizing Eq.~(\ref{ne22}), with respect to $t$ is not difficult; it can be readily shown that $t_{max}=\pi/\Omega$, a sort of ``$\pi$-pulse'' condition that will be consistent with $t<T$ if $T$ is chosen appropriately, specifically, provided that 
\begin{equation}
 -\sqrt{ \frac{64 \bar n^2 \Gamma_P^2}{\Gamma^2}-\pi^2} \le \frac{\Gamma T}{4} -  \frac{8 \bar n \Gamma_P}{\Gamma} \le \sqrt{ \frac{64 \bar n^2 \Gamma_P^2}{\Gamma^2}-\pi^2} 
 \label{ne23}
\end{equation}
Substituting $t=\pi/\Omega$ in (\ref{ne22}), we get the function
\begin{equation}
P_e = \frac{4\bar n \Gamma_P/T}{\Gamma^2+8\bar n \Gamma_P/T}\left(1+\exp\left[-\frac{3\pi\Gamma}{\sqrt{64 \bar n \Gamma_P/T-\Gamma^2}}\right]\right)
\label{ne24}
\end{equation}
This is a monotonically decreasing function of $T$, which will therefore be maximized by choosing the smallest value of $T$ that is compatible with the condition (\ref{ne23}).  
The result is a complicated function of $\bar n \Gamma_P/\Gamma$.  In the $\bar n = 1$ case, and for no external losses ($\Gamma_P=\Gamma$), it has the value $0.433$.

It turns out, however, that it is possible to do better than this, at least in the $\bar n =1$ case, by using a pulse that is \emph{shorter} than $\pi/\Omega$.  For such a pulse, the excitation probability grows monotonically and is maximum at end of the pulse ($t=T$), with the result
\begin{align}
P_e(T) = &\frac{4 \Gamma_P}{\Gamma^2 T + 8 \Gamma_P}\biggl(1 - e^{-3\Gamma T/4} \cr
&\quad\quad\times\left[\cos(\Omega T) + \frac{3\Gamma}{4 \Omega} \sin(\Omega T)\right]\biggr)
\label{ne25}
\end{align}
with $\Omega T = \frac 1 4 \sqrt{64 \Gamma_P T - \Gamma^2 T^2}$.  Equation (\ref{ne25}) depends fundamentally on two variables, which we can choose to be $\Gamma_P/\Gamma$ and $\Gamma T$.  For each value of the first one, we can numerically find the value of the second one that maximizes $P_e$.  For $\Gamma_P = \Gamma$ (no external losses), the optimal $T$ is found to be $T = 1.487/\Gamma$, and the maximum $P_e$ is
\begin{equation}
P_{e,max} = 0.482  \qquad \text{(square pulse)}
\end{equation}
For other values of the external losses, we get the results shown in Figure 2, which also includes the results of numerical calculations for other pulse shapes.  Note that the dependence on the external losses does not follow the simple dependance on $\Gamma_P/(\Gamma_P+\Gamma_B)$ that we obtained for single-photon Fock states in Section II.A above.

\begin{figure}
\begin{center}
\includegraphics[width=3.5in]{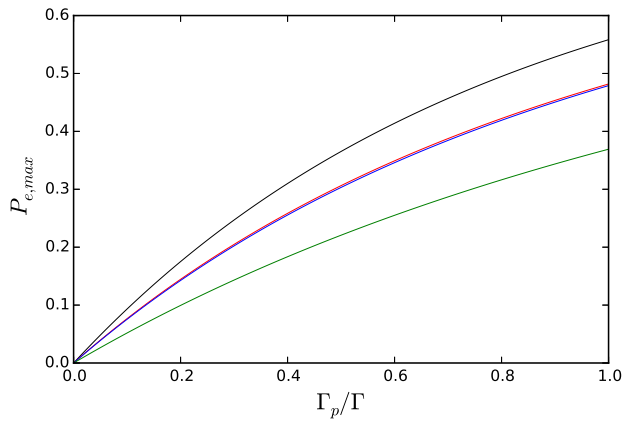}
\end{center}
\caption[example]
   { \label{fig:fig2}
Optimized excitation probability for various pulse shapes, as a function of the ratio $\Gamma_P/(\Gamma_P+\Gamma_B)$, for coherent states with $\bar n=1$.  From top to bottom, the curves are for a rising exponential pulse, a ``square'' pulse, a Gaussian pulse, and a decaying exponential pulse.  }
\end{figure}

\section{Multi-photon wavepackets, and asymptotic results}

\subsection{Coherent states}

\subsubsection{General results; square pulse}

When considering multiphoton excitation, especially in the large $\bar n$ limit, it makes more sense to think in terms of coherent states than Fock states, since multiphoton Fock states are notoriously difficult to produce.  Accordingly, we will consider coherent states first, in which case the basic equations to solve are just Eqs.~(\ref{ne19}), from Section II.  

As indicated earlier, Eqs.~(\ref{ne19}) cannot be solved exactly, to the best of our knowledge, except for a square pulse.  Approximate solutions, however, are possible in two opposite limits.  If the pulse is very long compared the overall decay time, $(\Gamma_P+\Gamma_B)^{-1}$, one can derive in a straightforward way an ``adiabatic solution,'' by formally setting the left-hand sides of Eqs.~(\ref{ne19}) equal to zero:
\begin{equation}
P_e(t) = \frac{4\bar n\Gamma_P}{(\Gamma_B+\Gamma_P)^2 + 8\bar n \Gamma_P f(t)^2}\,f(t)^2
\end{equation}
This more or less tracks the pulse, but it is always smaller than $1/2$, which is the value it approaches asymptotically as $\bar n\to \infty$.  This is the well-known phenomenon of ``bleaching'': a sufficiently long and intense classical pulse will drive the population inversion of a two-level medium to zero, so the medium becomes transparent.

We are interested here in a different regime, where we expect the excitation probability can be made to approach the instantaneous value of 1 for a sufficiently intense and short pulse.  Although, in general, this near-perfect excitation may be achieved for only a short time, one should note that, as long as the atomic levels are allowed to decay, any excitation we may produce will necessarily be transient. Whether it is useful or not depends on the timescales involved. 

As in the previous section, we are particularly interested in exploring the differences between pulse shapes, only this time we want to see how the pulse shape affects the rate at which $P_e$ approaches 1 as $\bar n$ increases.  We may conveniently start with the square pulse solution we derived above, Eq.~(\ref{ne22}), which has a local maximum at $t = \pi/\Omega = \pi T/\sqrt{4\bar n\Gamma_P T-(\Gamma T/4)^2}$, provided the condition (\ref{ne23}) is satisfied. For a sufficiently large $\bar n$, it is easy to see that this condition becomes
\begin{equation}
\frac{\pi^2 \Gamma }{4\bar n\Gamma_P} + O\left(\frac{1}{\bar n}\right)^3 \le \Gamma T \le \frac{64 \bar n\Gamma_P}{\Gamma} -\frac{\pi^2\Gamma }{4\bar n\Gamma_P} + O\left(\frac{1}{\bar n}\right)^3
\label{ne28}
\end{equation}
Also in the large $\bar n$ limit, Eq.~(\ref{ne24}) becomes
\begin{equation}
P_e \simeq 1 - \frac{3\pi\Gamma}{16}\sqrt{\frac{T}{\bar n \Gamma_P}} + \left(\frac{9\pi^2}{64}-\frac 1 2 \right) \frac{\Gamma^2 T}{4 \bar n\Gamma_P} +  O\left(\frac{1}{\bar n}\right)^{3/2}
\label{ne29}
\end{equation}
Note that, if we do not optimize the pulse duration $T$, Eq.~(\ref{ne29}) only approaches 1 as $1/\sqrt{\bar n}$.  On the other hand, if we substitute for $T$ the smallest value allowed by Eq.~(\ref{ne28}), namely, $\pi^2/4\bar n\Gamma_P$, we get a much more favorable scaling:
\begin{equation}
P_e \simeq 1 - \frac{3\pi^2\Gamma}{32 \bar n \Gamma_P} + \left(\frac{9\pi^2}{64}-\frac 1 2 \right) \left(\frac{\Gamma}{4 \bar n\Gamma_P}\right)^2 + O\left(\frac{1}{\bar n}\right)^3
\label{ne30}
\end{equation}
We should also verify that this is better than (or, as it turns out, equivalent to) the alternative we found for the $\bar n=1$ case in the previous section, namely, letting the maximum happen at the end of the pulse ($t=T$), in which case we need to optimize 
\begin{align}
P_e(T) = &\frac{4 \bar n \Gamma_P}{\Gamma^2 T + 8 \bar n \Gamma_P}\Biggl(1- e^{-3\Gamma T/4} \cr
&\quad\quad\times\left[\cos(\Omega T) + \frac{3\Gamma}{4 \Omega} \sin(\Omega T)\right]\Biggr)
\label{ne31}
\end{align}
with respect to $T$.  However, for large $\bar n$ it is clear that the prefactor approaches $1/2$, and the only way the term in parentheses can approach 2 is if $\cos(\Omega T) \simeq -1$.  This requires $T \simeq \pi/\Omega$, so at this point this approach reduces to the previous one, since there we started by imposing $t=\pi/\Omega$ and later choosing the lowest value of $T$ compatible with this condition, namely, $T=\pi/\Omega$.   

Finally, note that, in contrast to the single-photon case, in the large-$\bar n$ limit the optimal pulse duration (here $T=\pi^2/4\bar n\Gamma_P$) is, to lowest order in $1/\bar n$, independent of the external loss rate $\Gamma_B$.  We will find this to be the case for every other pulse shape, as well.

\subsubsection{Perturbation theory in the large $\bar n$ limit}

The above exactly-solvable case shows that, in order to get the first-order correction (deviation from unity), in $1/\bar n$, to $P_e$ it is enough to keep terms linear in $\Gamma$ (note that $\Gamma$ and $\Gamma_P$ are treated as completely independent variables here; $\Gamma_P$ characterizes the atom-field coupling, whereas $\Gamma$ quantifies the losses, or spontaneous decay rate).  It also suggests that, to the same order of accuracy, we may simply replace $\Omega = \sqrt{\Omega_0^2 + \Gamma^2/16}$ by $\Omega_0 = 2\sqrt{\bar n \Gamma_P/T}$, and assume that the maximum of $P_e$ happens at the time $t=\pi/\Omega_0$ where the $\pi$ pulse condition is satisfied in the absence of losses.  

This suggests a simple strategy to obtain the first-order correction for an arbitrary pulse shape, namely, to use perturbation theory.  Let $T$ be some parameter with the dimensions of time that characterizes the duration of the pulse, and let $g(t) = \sqrt T \, f(t)$ (so $g$ has the same shape as $f$ but is dimensionless).  Defining $\Omega_0 = 2\sqrt{\bar n \Gamma_P/T}$ as above and the dimensionless time $\tau = \Omega_0 t$, we can rewrite the system (\ref{ne19}) as
\begin{align}
\begin{split}
& \frac{d x}{d\tau} = -\frac{\epsilon}{2}\,x + g(\tau) y - g(\tau)
\cr
& \frac{d y}{d\tau} = -\epsilon\, y -g(\tau) x
\end{split}
\label{ne32}
\end{align}
where $x \equiv {\Sigma}$, $y = 2 P_{e}$, and $\epsilon = \Gamma/\Omega_0$.  We can then expand $x(t) = x^{(0)}(t) + \epsilon x^{(1)}(t)+\ldots$, $y(t) = y^{(0)}(t) + \epsilon y^{(1)}(t)+\ldots$, and substitute in (\ref{ne32}).  The lowest-order equation
\begin{align}
\begin{split}
& \frac{d x^{(0)}}{d\tau} =  g(\tau)\, y^{(0)} - g(\tau)
\cr
& \frac{d y^{(0)}}{d\tau} = -g(\tau)\, x^{(0)}
\end{split}
\label{ne33}
\end{align}
is immediately solved by 
\begin{align}
x^{(0)}(\tau) &= -\sin[\theta(\tau)] \cr
y^{(0)}(\tau) &= 1-\cos[\theta(\tau)] 
\label{ne34}
\end{align}
with
\begin{equation}
\theta(\tau) = \int_{-\infty}^\tau g(\tau^\prime)\,d\tau^\prime
\label{ne35}
\end{equation}
and, as expected, this gives unit excitation probability when $\theta = \pi$.  The next-order correction must satisfy
\begin{align}
\begin{split}
& \frac{d x^{(1)}}{d\tau} = g(\tau)\, y^{(1)} -\frac{\epsilon}{2}\,x^{(0)}(\tau)
\cr
& \frac{d y^{(1)}}{d\tau} =  -g(\tau)\, x^{(1)} -\epsilon\, y^{(0)}(\tau)
\end{split}
\label{ne36}
\end{align}
Again, changing to the variable $\theta$ turns this into a simple driven harmonic oscillator problem, with the formal solution for $y^{(1)}(\theta)$:
\begin{widetext}
\begin{equation}
y^{(1)}(\theta) = -\epsilon\int_0^\theta \frac{d\tau}{d\theta^\prime}\left(y^{(0)}(\theta^\prime)\cos(\theta-\theta^\prime) - \frac 1 2 x^{(0)}(\theta^\prime)\sin(\theta-\theta^\prime) \right) d\theta^\prime
\label{ne37}
\end{equation}
At this point, the only remaining difficulty may be to express $d\tau/d\theta = 1/g(\tau(\theta))$ as a function of $\theta$, since the inversion of Eq.~(\ref{ne35}) may be a nontrivial problem.  Alternatively, note that the whole expression (\ref{ne37}) can be rewritten explicitly as an integral over $\tau$. For the moment, though, we will continue to use $\theta$ because it allows us to express the correction to $P_e$ at the expected maximum, $\theta = \pi$, in the following very compact form (using Eqs.~(\ref{ne34}), (\ref{ne35})):
\begin{align}
1-P_e \Bigr|_{\theta=\pi} &= \frac{\epsilon}{2}\int_0^\pi \frac{1}{g(\tau(\theta^\prime))}\left(-\cos(\theta^\prime)\left[1-\cos(\theta^\prime)\right] + \frac 1 2 \sin^2(\theta^\prime) \right) d\theta^\prime \cr
&= \epsilon \int_0^\pi \frac{\sin^4(\theta/2)}{g(\tau(\theta))} d\theta
\label{ne38}
\end{align}
\end{widetext}
Examples of the use of Eq.~(\ref{ne38}) follow.

\subsubsection{Decaying exponential pulse}
Consider a decaying exponential pulse, $f(t) = e^{-t/T}\sqrt{2/T}$ for $t\ge 0$, and zero for $t<0$.  Then $g(\tau) = \sqrt 2\, e^{-\tau/\Omega_0 T}$, $\theta = \sqrt 2\, \Omega_0 T (1-e^{-\tau/\Omega_0 T})$, and $g(\tau(\theta)) = \sqrt 2[1-\theta/(\sqrt 2\Omega_0 T)]$.  The result is then
\begin{equation}
1-P_e \Bigr|_{\theta=\pi} = \Gamma T  \int_0^\pi \frac{\sin^4(\theta/2)}{2\sqrt{2\bar n\Gamma_P T}-\theta}\, d\theta
\label{ne39}
\end{equation}
It is now an easy matter to minimize this, numerically, with respect to $T$.  The minimum is obtained when $T=3.347/\bar n \Gamma_P$, and the result is then
\begin{equation}
P_e  \simeq 1 - 1.47895 \frac{\Gamma}{\bar n \Gamma_P}
\label{ne40}
\end{equation}
This is clearly less favorable than the square-pulse scaling, Eq.~(\ref{ne30}), since the prefactor $3\pi^2/32 \simeq 0.9253$.  Again, note that if one were simply to increase $\bar n$ in Eq.~(\ref{ne39}), leaving $T$ constant, the excitation probability would only approach 1 as $1/\sqrt{\bar n}$, which is to say, much more slowly.

\subsubsection{Rising exponential pulse}

Let now $f(t) = e^{t/T}\sqrt{2/T}$ for $t\le 0$, and zero for $t>0$.  Then $g(\tau) = \sqrt 2\, e^{\tau/\Omega_0 T}$ (for $\tau\le 0$), and $\theta = \sqrt 2\, \Omega_0 T e^{\tau/\Omega_0 T}$, so $g(\tau(\theta)) = \theta/\Omega_0 T$, and we find
\begin{equation}
1-P_e \Bigr|_{\theta=\pi} = \Gamma T  \int_0^\pi \frac{\sin^4(\theta/2)}{\theta}\, d\theta = 0.519432\, \Gamma T
\label{ne41}
\end{equation}
This is, at first sight, a somewhat surprising result, in that it looks like it can be made arbitrarily small simply by reducing $T$, but recall that in order for Eq.~(\ref{ne38}) to be applicable, it must be possible for $\theta$ to reach the value of $\pi$, so we need to have $\sqrt 2\, \Omega_0 T \ge \pi$.  Since $\Omega_0 = 2\sqrt{\bar n \Gamma_P/T}$, this leads to the condition 
\begin{equation}
2\sqrt{2\bar n \Gamma_P T} \ge \pi
\label{ne42}
\end{equation}
Taking the smallest $T$ compatible with Eq.~(\ref{ne42}), namely, $T_{opt} = \pi^2/8\bar n\Gamma_P$, and substituting in (\ref{ne41}), we obtain
\begin{equation}
P_e  = 1 - 0.519432\, \frac{\Gamma \pi^2}{8 \bar n \Gamma_P} = 1 - 0.640824 \frac{\Gamma}{\bar n \Gamma_P}
\label{ne41}
\end{equation}
This is better than the square pulse, and much better than the decaying exponential, requiring less than half the photons to reach the same value of $P_e$.

\subsubsection{Gaussian pulse}
Finally, consider a Gaussian pulse of the form $f(t) = e^{-t^2/T^2}/\sqrt{T\sqrt{\pi/2}}$.  We now have
\begin{equation}
g(\tau) = \left(\frac{2}{\pi}\right)^{1/4} e^{-\tau^2/(\Omega_0 T)^2}
\end{equation}
\begin{equation}
\theta(\tau) = \int_{-\infty}^\tau g(\tau^\prime)\, d\tau^\prime = \frac 1 2 \Omega_0 T (2\pi)^{1/4} \left[1+\text{erf}\left(\frac{\tau}{\Omega_0 T}\right)\right]
\end{equation}
\begin{align}
1-P_e \Bigr|_{\theta=\pi} = &\frac{\Gamma}{\Omega_0} \int_0^\pi\exp\left[\text{InverseErf}\,^2\left(\frac{2\theta}{\Omega_0 T}\,\frac{1}{(2\pi)^{1/4}} - 1\right)\right]\cr
&\times\sin^4\left(\frac\theta 2\right)\,d\theta
\label{ne46}
\end{align}
where ``InverseErf'' is the inverse of the error function, available in packages such as \emph{Mathematica}. One now has to (numerically) minimize (\ref{ne46}) with respect to $T$, keeping in mind that $\Omega_0$ itself depends on $T$.  In practice, it is easiest to introduce a parameter $a= \Omega_0 T$ in terms of which $T= a^2/4\bar n \Gamma_P$, and the prefactor $\Gamma/\Omega_0 = \Gamma T/a = a \Gamma/4\bar n \Gamma_P$, and minimize the resulting expression with respect to $a$.  One then finds that 
\begin{equation}
T_{opt} = \frac{1.45009}{\bar n \Gamma_P}
\end{equation}
and
\begin{equation}
P_e = 1 - 0.91597\,\frac{\Gamma}{\bar n \Gamma_P}
\label{ne48}
\end{equation}
very close to the square pulse result, Eq.~(\ref{ne30}), which is $\simeq 1 - 0.9253 \,{\Gamma/\bar n \Gamma_P}$ to lowest order.  All of the above results are summarized graphically (for the lossless case) in Fig.~3.
\begin{figure}
\begin{center}
\includegraphics[width=3.7in]{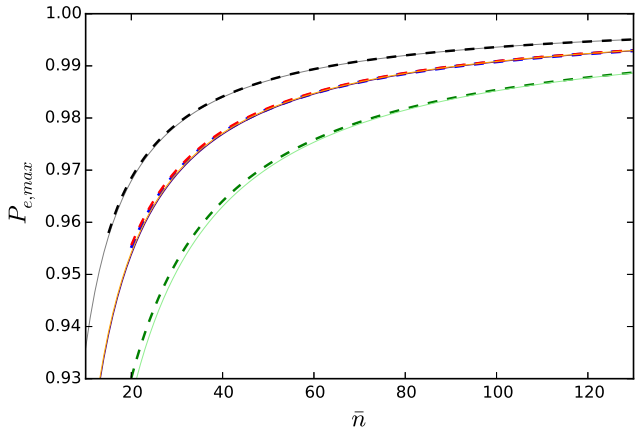}
\end{center}
\caption[example]
   { \label{fig:fig3}
Optimized excitation probability for multiphoton coherent states, for various pulse shapes, as a function of the average photon number $\bar n$, in the absence of external losses ($\Gamma_B=0$).  From top to bottom, the curves are for a rising exponential pulse, a ``square'' pulse, a Gaussian pulse (these two are virtually indistinguishable on this scale), and a decaying exponential pulse.  The solid lines show the analytical approximation, and the dashed lines the results of numerical calculations.}
\end{figure}

\subsection{Fock states}

Although multiphoton Fock states are very difficult to prepare, we wish to cover this case here for completeness, since it also turns out to be analytically tractable in the large $N$ limit. 

Generalizing the model in \cite{scarani2} to include external losses, we find that the excitation probability for $N$ photons can be obtained by integrating the following system of  $2N$ coupled differential equations: 
\begin{align}
\dot{P}_{e,n} &=-\Gamma P_{e,n}-\sqrt{\Gamma_{P} n}f(t)\;\Sigma_{n-1}
\cr
\dot{\Sigma}_{n-1} &=-\dfrac{\Gamma}{2}\;\Sigma_{n-1}+4\sqrt{\Gamma_{P} n}f(t)\;P_{e,n-1}-2\sqrt{\Gamma_{P} n}f(t) \cr
\label{e9}
\end{align}
where the index $n$ runs from $1$ to $N$.  (For an alternative formalism to deal with this problem, see the work of Baragiola et al. \cite{combes}.) Let $f(t) = 1/\sqrt T$, $0<t<T$.  In this case, and for a small number of photons, one could easily integrate Eq.~(\ref{e9}) recursively, by hand, and obtain the excitation probability. However, this becomes impractical for a significantly large number of photons. We therefore resort to the same kind of perturbation theory we used above for the coherent-state pulse.  

Letting $g(t)=\sqrt T\,f(t)$, $\Omega_0 = 2\sqrt{N\Gamma_P T}$, $\tau = \Omega_0 t$, we find the system (\ref{e9}) can be written as
\begin{align}
\frac{d}{d\tau}\,y_n &= -\epsilon y_n - g(\tau)\sqrt{\frac n N}\, x_{n-1} \cr
\frac{d}{d\tau}\,x_{n-1} &= -\frac \epsilon 2\, x_{n-1} + g(\tau)\sqrt{\frac{n}{N}}\, y_{n-1} - g(\tau) \sqrt{\frac{n}{N}}\cr
\label{e50}
\end{align}
where $x_n = \Sigma_n$, $y_n = 2 P_n$, and $\epsilon = \Gamma/\Omega_0$.

Introducing a perturbative solution of the form $x_n(t) = x_n^{(0)} + \epsilon x_n^{(1)} +\ldots$, $y_n(t) = y_n^{(0)} + \epsilon y_n^{(1)} +\ldots$, one can show recursively that the zero-th order solution has the form
\begin{align}
y_n^{(0)}(\theta) &=1 -  \mathbf{ _{1}F_{1}} \left(-n,\frac 1 2, \frac{\theta^2}{4 N}\right) \cr
x_{n-1}^{(0)}(\theta) &= -\theta \sqrt\frac{n}{N}\, \mathbf{ _{1}F_{1}} \left(-n+1,\frac 3 2, \frac{\theta^2}{4 N}\right) 
\label{e51}
\end{align}
in terms of the variable $\theta$ introduced as in Eq.~(\ref{ne35}), and the confluent hypergeometric function $\mathbf{ _{1}F_{1}}$.  This is to be compared directly to the result (\ref{ne34}) for the coherent state case, with $n=N$.  Indeed, in the large $N$ limit we find \cite{abr1}
\begin{align}
 \mathbf{ _{1}F_{1}} \left(-N,\frac 1 2, \frac{\theta^2}{4 N}\right) &\simeq e^{\theta^2/8N} \cos\theta \cr 
  \mathbf{ _{1}F_{1}} \left(-N,\frac 3 2, \frac{\theta^2}{4 N}\right) &\simeq \frac 1 \theta\,e^{\theta^2/8N} \sin\theta 
\label{ne52}
\end{align}
which shows that, as in the coherent-state case, the excitation probability, $y_N/2$, will be maximum around $\theta = \pi$. Note, however, that since we ultimately want an expression for $P_e$ that is correct to order $1/N$, we cannot neglect the exponential term in (\ref{ne52}) completely.  Rather, we have to say that, to lowest order in $\epsilon$, the excitation probability, $P_e^{(0)}$ is given by
\begin{align}
P_e^{(0)} &= \frac 1 2(1- e^{\theta^2/8N} \cos\theta) \cr
P_{e,max}^{(0)} &\simeq 1 + \frac{\pi^2}{16 N}
\label{ne53}
\end{align}
Of course, an excitation probability greater than 1 is unphysical, but this is ultimately due to the fact that the zero-th order in $\epsilon$ is also unphysical: since $\epsilon = (\Gamma_P+\Gamma_B)/\Omega_0$ includes the coupling to the atom $\Gamma_P$, it can never be strictly zero.  As we shall see below, the terms coming from the first order correction will ensure that $P_e$ is always less than 1, to first order in $1/N$.

This first-order correction is formally given by 
\begin{widetext}
\begin{align}
y_N^{(1)}(\theta) = -\epsilon\int_0^\theta \frac{d\tau}{d\theta^\prime}&\Biggl\{
\sum_{n=0}^{N-1} \frac{(-1)^n N!}{(2n)! N^n (N-n)!}\,(\theta-\theta^\prime)^{2n}\left[1 -  \mathbf{ _{1}F_{1}} \left(-N+n,\frac 1 2, \frac{{\theta^\prime}^2}{4 N}\right)\right] \cr
&-\frac 1 2 \sum_{n=1}^{N} \frac{(-1)^{n-1} (N-1)! }{(2n-1)! N^{n-1} (N-n)!}\,{(\theta-\theta^\prime)^{2n-1}}{\theta^\prime} \mathbf{ _{1}F_{1}} \left(-N+n,\frac 3 2, \frac{{\theta^\prime}^2}{4 N}\right) \Biggr\} d\theta^\prime 
\label{ne54}
\end{align}
\end{widetext}
where, again, we wish to emphasize the similarity with the corresponding coherent-state result (\ref{ne37}).  In fact, it is straightforward to show numerically that, for finite $\theta$ (in particular, $\theta\simeq \pi$), the term in curly braces in Eq.~(\ref{ne54}) approaches $\cos(\theta-\theta^\prime)(1-\cos\theta^\prime -\frac 12 \sin(\theta-\theta^\prime)\sin\theta^\prime)$, as $N\to\infty$.  Qualitatively, this may be understood from the fact that, for large $N$ and small $n$, the difference between $N$ and $N-n$ in the first argument of the hypergeometric functions can be neglected, and for large $n$ the prefactors go to zero very fast, so the terms where the difference between $N$ and $N-n$ is substantial are strongly suppressed.  Setting, then $N-n \simeq N$ in the argument of the hypergeometric functions, the sums can be carried out, with the result that the first one equals $\mathbf{ _{1}F_{1}} \left(-N,\frac 1 2, (\theta-\theta^\prime)^2/{4 N}\right)$ (up to terms that are negligible for large $N$), and the second one equals  $(\theta-\theta^\prime)\mathbf{ _{1}F_{1}} \left(-N+n,\frac 3 2,(\theta-\theta^\prime)^2/4 N\right)$.  One can then use the results (\ref{ne52}) to show the (asymptotic) identity between (\ref{ne54}) and the coherent-state result  (\ref{ne34}), (\ref{ne37}).  In particular, for $\theta = \pi$, we have then (making use of (\ref{ne53}))
\begin{equation}
1-P_e\Bigr|_{\theta=\pi} = -\frac{\pi^2}{16 N} + \epsilon \int_0^\pi \frac{\sin^4(\theta/2)}{g(\tau(\theta))} d\theta
\label{ne55}
\end{equation}
which means that the optimal pulse bandwidth (to minimize the second term on the right-hand side of (\ref{ne55})) will be exactly the same as for the coherent-state case (only with $\bar n$ replaced by $N$), and the maximum excitation probability will also be the same, plus the $\pi^2/16N$ term.  Explicitly, we get
\begin{align}
P_e &= 1 -\frac{\pi^2}{32 N} -\frac{3\pi^2 \Gamma_B}{32 N \Gamma_P} \qquad\text{(square pulse)}\cr
P_e &= 1 -\frac{0.8621}{N} -\frac{1.47895\Gamma_B}{N \Gamma_P} \qquad\text{(decaying exponential)}\cr
P_e &= 1 -\frac{0.02397}{N} -\frac{0.64082\Gamma_B}{N \Gamma_P} \qquad\text{(rising exponential)}\cr
P_e &= 1 -\frac{0.29912}{N} -\frac{0.91597\Gamma_B}{N \Gamma_P} \qquad\text{(Gaussian)}
\label{ne56}
\end{align}
where we have separated the contribution of the external losses $\Gamma_B$ explicitly, to show that $P_e$ is indeed in all the cases lower than 1. (Note that this $\Gamma_B$ contribution is the same for Fock states as for coherent states.) We find that the rising exponential pulse in the $\Gamma_B = 0$ case is now more than an order of magnitude better than all the other pulses.  These results are plotted (for the $\Gamma_B=0$ case) in Figure 4.
\begin{figure}
\begin{center}
\includegraphics[width=3.7in]{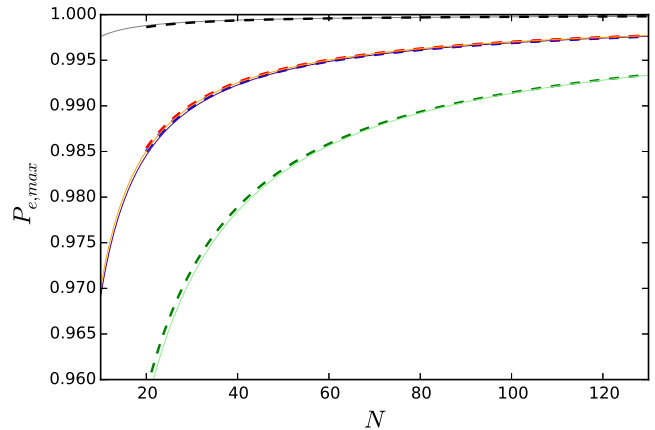}
\end{center}
\caption[example]
   { \label{fig:fig4}
Optimized excitation probability for multiphoton Fock states, for various pulse shapes, as a function of the photon number $N$, in the absence of external losses ($\Gamma_B=0$).  From top to bottom, the curves are for a rising exponential pulse, a ``square'' pulse, a Gaussian pulse (these two are virtually indistinguishable on this scale), and a decaying exponential pulse. The solid lines show the analytical approximation, and the dashed lines the results of numerical calculations. }
\end{figure}

The last of the equations (\ref{ne56}) should be compared (in the $\Gamma_B=0$ case) to the result obtained for Fock states by Baragiola et al., $P_e^N = 1-0.269 N^{-0.973}$ (see caption to Figure 3 of \cite{combes}).  The two expressions yield very similar results for $N=40$, which was the largest value of $N$ considered in \cite{combes}, and our numerical calculations show that ours is a better fit for large $N$, as is to be expected.

\section{Conclusions}

We have considered theoretically the maximum excitation probability for a two-level system interacting with a quantum field, in the presence of external losses (or non-perfect coupling), in two complementary limits: when the incident field contains a single photon (on average), and asymptotically, when the number of photons is very large.  In both cases we find that $P_e$ depends strongly on the temporal profile of the pulse, and that for each pulse shape it is essential to optimize the pulse duration (or bandwidth) in order to make $P_e$ as large as possible.  In particular, for the single-photon (Fock state) case, we find that, if the external losses are characterized by the decay rate $\Gamma_B$, the optimum bandwidth depends on $\Gamma_B$, and with this optimization the maximum $P_e$ is just equal to the lossless result multiplied by the ratio $\Gamma_P/(\Gamma_P+\Gamma_B)$.  For a coherent state with $\bar n = 1$, we have presented numerical results showing that this simple scaling does not apply.

For the multiphoton case, when the field is in a coherent state, we have derived an expression that allows one to evaluate the leading term in the expansion of $P_e$ in powers of $1/\bar n$ for a pulse of arbitrary temporal profile.  We find in this case that the optimum pulse duration does not depend on the external losses, and what is typically required is for $T$ to scale as $\alpha/\bar n \Gamma_P$, where the constant $\alpha$ depends on the pulse shape.  When this is done, one finds that $P_e$ approaches 1 as $P_e \simeq 1 - (\beta/\bar n)(1+\Gamma_B/\Gamma_P)$, where again the constant $\beta$ depends on the pulse shape.  When the pulse duration is not optimized, one typically finds a much less favorable scaling with $\bar n$.  We also find that the constant $\beta$ can vary by a factor of 2 or more across different pulses, and in terms of their effectiveness the various shapes that we have studied rank in roughly the same order in the large $\bar n$ as in the single-photon regime, with the rising exponential profile being the best, the decaying exponential the worst, and the square and Gaussian profiles being in the middle and very close to each other.

For multiphoton Fock states, we have shown that the optimal pulse bandwidth and the time when the excitation peaks are (in the asymptotic, large $N$ limit) the same as for the corresponding coherent state with $\bar n = N$, and the optimized excitation probability is the same plus $\pi^2/16 N$, regardless of the pulse shape or the level of external losses.

\end{document}